# Dynamically encircling exceptional points in a waveguide: asymmetric mode switching from the breakdown of adiabaticity

Jörg Doppler[1], Alexei A. Mailybaev[2], Julian Böhm[3], Ulrich Kuhl[3], Adrian Girschik[1], Florian Libisch[1], Thomas J. Milburn[4], Peter Rabl[4], Nimrod Moiseyev[5], and Stefan Rotter[1]

**Affiliations:**

[1]Institute for Theoretical Physics, Vienna University of Technology (TU Wien), Vienna, A-1040, Austria, EU

[2]Instituto Nacional de Matemática Pura e Aplicada - IMPA, 22460-320 Rio de Janeiro, RJ, Brazil

[3]Laboratoire de Physique de la Matière Condensée, CNRS UMR 7336, Université Nice Sophia Antipolis, 06108 Nice, France, EU

[4]Institute of Atomic and Subatomic Physics, Vienna University of Technology (TU Wien), Vienna, A-1040, Austria, EU

[5]Schulich Faculty of Chemistry and Faculty of Physics, Technion - Israel Institute of Technology, Haifa, 32000, Israel

**Physical systems with loss or gain feature resonant modes that are decaying or growing exponentially with time. Whenever two such modes coalesce both in their resonant frequency and their rate of decay or growth, a so-called "exceptional point" occurs, around which many fascinating phenomena have recently been reported to arise[1–6]. Particularly intriguing behavior is predicted to appear when encircling an exceptional point sufficiently slowly[7,8], like a state-flip or the accumulation of a geometric phase[9,10]. Experiments dedicated to this issue could already successfully explore the topological structure of exceptional points[11–13], but a full dynamical encircling and the breakdown of adiabaticity inevitably associated with it[14–21] remained out of reach of any measurement so far. Here we**

**demonstrate that a dynamical encircling of an exceptional point can be mapped onto the problem of scattering through a two-mode waveguide, which allows us for the first time to access the elusive effects occurring in this context. Specifically, we present experimental results from a waveguide structure that steers incoming waves around an exceptional point during the transmission process. In this way mode transitions are induced that make this device perfectly suited as a robust and asymmetric switch between different waveguide modes. Our work opens up new and exciting avenues to explore exceptional point physics at the crossroads between fundamental research and practical applications.**

Exceptional points (EPs), also called non-Hermitian degeneracies or branch points, have recently attracted considerable attention. This is because these singular points have turned out to be at the origin of many counter-intuitive phenomena appearing in physical systems that experience gain or loss[1–6]. Such external influences on a system require a non-Hermitian description that incorporates non-conservation of energy resulting from an external in- or output. Rather than being merely a perturbative correction, gain and loss can entirely turn the behavior of a system upside down when approaching an EP. Consider here, e.g., the recent demonstrations of unidirectional invisibility[22–25], loss-induced suppression and revival of lasing[26,27], and single-mode lasers with gain and loss[28,29], all of which were realized at or close to an EP. While these studies already nicely demonstrate the potential of EPs for novel effects and devices, their full capability can be brought to bear when the EP is not just approached or swept across, but parametrically encircled[7,8].

Originally, it was believed that a slow encircling of an EP would result in an adiabatic evolution of states and a corresponding state flip[9], but more recent work has rigorously shown that the very same non-Hermitian components necessary for the observation of an EP actually prevent an application of the adiabatic theorem[14–16,18–21]. Instead, the non-adiabatic terms lead to a chiral behavior, in the sense that encircling an EP in a clockwise and a counter-clockwise direction results in different final states[14,18,21]. While this fascinating feature has great potential for quantum control and switching protocols, it has so far defied any experimental realization. This is because to observe the non-adiabatic contributions requires a fully dynamical encircling of the EP that goes beyond the quasi-static experiments reported so far[11–13]. A dynamically resolved experiment is, however, extremely challenging, because of the required precise control of exponentially amplified or damped resonant modes which meet at the EP, and which must be decoupled from all other modes present in a system. Proposals to overcome this problem have meanwhile been put forward, such as to map the dynamical encircling of an EP to the polarization evolution in a stratified non-transparent medium[15], but the involved implementation requirements prevented an experimental realization also for this case. Here, we overcome such difficulties by demonstrating that waveguides with two transverse modes can be suitably engineered such that the transmission through them is equivalent to a slow dynamical encircling of an EP. In this way we make the recently discussed dynamical features of EPs directly accessible through established waveguide technology as used for the transmission of sound, light, micro- and matter waves.

An EP arises when an open system described by a Schrödinger-type equation $i\partial_t \psi = H\psi$ features two resonant modes that coalesce. Such a scenario can conveniently be captured already by the following non-Hermitian 2×2 Hamiltonian,

$$H = \begin{pmatrix} \delta - i\gamma_1/2 & g \\ g & -i\gamma_2/2 \end{pmatrix}, \tag{1}$$

where $g$ denotes the coupling and $\delta$ the detuning, $\gamma_1$ and $\gamma_2$ are the respective loss rates of the two relevant modes. At the specific parameter configuration $\delta_{EP} = 0$ and $g_{EP} = |\gamma_1 - \gamma_2|/4$, both the eigenvalues and eigenvectors of this Hamiltonian coalesce, which is the hallmark of the EP. As shown in Fig. 1, the vicinity of this point exhibits a characteristic structure of a self-intersecting Riemann surface. The EP marks the branch point (centered in each figure) at which the Riemann surface splits. It is this topological structure that allows one to encircle the EP such that the two eigenmodes interchange: For such a state-flip two system parameters need to be changed in time $t$ (e.g., the coupling $g = g(t)$ and the detuning $\delta = \delta(t)$) along a closed loop in parameter space around the EP. This system evolution is described by the now time-dependent Hamiltonian (1) in the corresponding Schrödinger-type equation $i\partial_t \psi(t) = H(t)\psi(t)$. If the system dynamics is fully adiabatic, a flip between the two states is realized upon encircling the EP such that the lower state becomes the upper one (Fig. 1a left). As was found only recently[14], however, contributions due to the breakdown of adiabaticity in non-Hermitian systems always enter dominantly whenever both encircling directions are considered: In the case above, traversing the same parameter loop in the opposite direction thus leads to the situation that the lower state returns to itself rather than to the upper state (Fig. 1b left). This enforces an overall asymmetric behavior such that the state that is selected at the end of a loop depends only on the loop's encircling direction, but not on its starting point – compare here Fig. 1a and 1b for a counter-clockwise and clockwise encircling, respectively. On a very fundamental level, these features are connected with the Stokes phenomenon of asymptotics[15,16,21] as well as with the theory of singular perturbations and stability loss delay[21].

To observe this behavior in a realistic environment, we now map the Hamiltonian in Eq. (1) onto the problem of microwave transmission through a smoothly deformed metallic waveguide in the presence of absorption (see Fig. 1c). The waveguide is extended along the $x$-axis and we restrict the following discussion to a single transverse dimension $y$. Within this framework, the parametrical encircling of the EP from the 2×2 model shown above translates to a slow variation of a periodic boundary modulation along the waveguide. Directly at the EP, both the Bloch wavenumber $K$ and the Bloch modes $\Lambda$ of the electric field distribution $\phi(x,y) = \Lambda(x,y)\, e^{iKx}$ coalesce. More specifically, the harmonic solutions $\varphi(x,y,t) = \phi(x,y)e^{-i\omega t}$ for fields oscillating with frequency $\omega$ obey the Helmholtz equation

$$\Delta \phi(x,y) + V(x,y)\phi(x,y) = 0, \qquad (2)$$

where $\Delta$ is the Laplace operator in 2D, $V(x,y) = \varepsilon(x,y)\,\omega^2/c^2$ is a complex potential proportional to the dielectric constant $\varepsilon$ and $c$ is the speed of light. For a straight rectangular waveguide with a fixed width $W$ in $y$-direction the solutions of Eq. (2) in the absence of losses are $\phi_n(x,y) = u_n(y)e^{ik_n x}$ with transverse mode functions $u_n(y) = \sin(n\pi y/W)$ and wavevectors $k_n = \sqrt{\omega^2/c^2 - n^2\pi^2/W^2}$. By choosing an appropriate input frequency $\omega$, the transmission problem can naturally be reduced to only two propagating modes $n = 1,2$. To implement a controlled coupling between these modes, we consider a waveguide subject to a boundary modulation $\xi(x) = \sigma \sin(k_b x)$, as shown in Fig. 1c. By choosing the boundary wavenumber $k_b = k_1 - k_2 + \delta$, where $|\delta| \ll k_b$, near resonant scattering between the otherwise very different modes $\phi_1$ and $\phi_2$ occurs. The full solution for the propagating field can be written in the form

$$\phi(x,y) = \frac{c_1(x)}{\sqrt{k_1}} \sin\left(\frac{\pi}{W}y\right) + \frac{c_2(x)}{\sqrt{k_2}} \sin\left(\frac{2\pi}{W}y\right) e^{-ik_b x} \ . \tag{3}$$

Employing a Floquet-Bloch ansatz, we obtain a Schrödinger-type equation for the slowly varying modal amplitudes $\psi(x) = \bigl(c_1(x), c_2(x)\bigr)^T$,

$$i\partial_x \begin{pmatrix} c_1(x) \\ c_2(x) \end{pmatrix} = \begin{pmatrix} \delta(x) - i\gamma_1/2 & g(x) \\ g(x) & -i\gamma_2/2 \end{pmatrix} \begin{pmatrix} c_1(x) \\ c_2(x) \end{pmatrix}. \tag{4}$$

(See the Supplementary Information for a more detailed derivation verified by exact numerical simulations.) The slow variation of $\delta = \delta(x)$ and $g = g(x) \propto \sigma(x)$ in Hamiltonian (1) is then directly implemented in the waveguide through a smooth variation of the modulation potential $V(x,y)$, which leaves the validity of Eqs. (3) and (4) intact. Finally, due to the even and odd symmetry of $u_1(y)$ and $u_2(y)$, an absorbing material placed close to the center of the waveguide gives rise to the required losses $\gamma_1 \gg \gamma_2$. With the above, all parameters in the non-Hermitian Hamiltonian $H$ in Eq. (1) are determined. However, instead of governing the temporal dynamics (in time), $H$ determines here the mode propagation in the longitudinal direction $x$. Correspondingly, the requirement of encircling the EP slowly (in time $t$) is transferred here to a slow variation of the boundary parameters along the propagation direction $x$ (see Fig. 1c). Quite remarkably, a right and left injection into the waveguide corresponds to a clockwise and counter-clockwise encircling direction of the EP, respectively, yielding a specific and different output mode depending only on the side from which the waves are injected.

First numerical results following this procedure are shown in Fig. 2, where we rely on a parametrization of the waveguide modulation envelope, $\sigma(x) = (\sigma_0/2)(1 - \cos 2\pi x/L)$, that is now restricted to a finite region ($x \in [0, L]$) and perfectly connects to flat semi-infinite waveguides outside this region. Deviating from what is shown in Fig. 1a,b, we also choose the detuning $\delta$ to be linear in $x$, $\delta(x) = \delta_0(2x/L - 1) + \rho$, which, together with $\sigma(x)$ from above, still describes a loop around the EP, since the endpoints of this parameter-trajectory correspond to identical waveguide configurations (see the Supplementary Information for details). By implementing these design considerations in a waveguide with uniform (bulk) loss in transverse direction, the desired asymmetric switching of modes is, indeed, fully realized: Either mode entering from the left (Fig. 2a,b) is scattered into the first mode at the right exit lead. By contrast, any mode injected from the right side of the waveguide yields the second mode at the left exit lead (Fig. 2c,d). On the downside, however, the large overall loss both states have to acquire in order to manifest this asymmetry considerably deteriorates the quality of this switching mechanism. Additionally, the requirement of slow encircling translates into a long and bulky device with many boundary oscillations. To overcome both of these obstacles, we devised the following two strategies: Firstly, we designed the absorption in the waveguide to follow a spatial pattern that minimizes (maximizes) the dissipation for the mode featuring the adiabatic (non-adiabatic) transition, while leaving the topology of the loop around the EP intact (see the Supplementary Information); quite remarkably, no matter which spatial profile we choose for the absorber, the reciprocity principle ensures that our design works for both transmission directions equivalently. Secondly, we employed a combination of quasi-Newton methods with stochastic algorithms to decrease the system length, resulting in a length-to-width ratio reduced by a factor of four as compared to the devices shown in Fig. 2. In this optimization, we tuned the parameters $\sigma_0$, $\delta_0$ and $\rho$ such as to reduce the waveguide length while making sure that the resulting device still maintains the frequency robustness inherent in our design principle (see Supplementary Fig. 8 for this efficient device geometry and the corresponding numerical results).

To demonstrate its potential for real-world applications, we provide here the first experimental realization of the above protocol, implemented in a surface-modulated microwave setup following the proposed efficient design (Fig. 3a,b). Measuring the modal transmission intensities $T_{nm}$ from mode $n$ into mode $m$ as a function of the input signal frequency, we unambiguously confirm the asymmetric switching effect (see Fig. 3c): An arbitrary combination of modes injected from the left side of the waveguide is transmitted into the first mode when arriving at the exit lead on the right ($T_{11}$ and $T_{21}$ dominate the transmission of the first and second mode, respectively, with transmission intensity ratios $T_{11}/T_{12} = 20.6$ and $T_{21}/T_{22} = 23.0$). At the same time, the second mode is produced on the left for injection from the right ($T'_{12}$ and $T'_{22}$ dominate the respective modal transmission, where the primed quantities are those for injection from the right, with ratios $T'_{12}/T'_{11} = 463.4 \approx T_{21}/T_{11} = 488.6$ and $T'_{22}/T'_{21} = 425.9 \approx T_{22}/T_{12} = 438.4$). Note that the slight violation of the reciprocity property $T'_{nm} = T_{mn}$ observed in the experiment (see Fig. 3c) is due to the magnetized absorber material (see details in the Methods section) that is needed to obtain a sufficiently strong absorption in the corresponding frequency range (without the absorber, the experiment is fully reciprocal). This small non-reciprocity is, however, not essential for the operation of our device, since the respective intensity ratios are approximately the same for both injection directions. Most importantly, the experimental data proves the very strong robustness of these transmission values with respect to variations of the input frequency - a broad-band feature that is a direct consequence of our design principle, which ensures operability also in the presence of small variations of the waveguide parameters. The shortened device for which the length-to-width ratio is now $L/W = 25$ also vastly outperforms the longer device in Fig. 2 (for which $L/W = 100$), not only in terms of length-to-width ratio, but also in terms of the output intensity which is here increased by six orders of magnitude.

In summary, our work constitutes the first experimental encircling of an exceptional point that stays faithful to the full dynamical and non-adiabatic behavior occurring in this context. In this way, we have devised a notably platform-independent approach to mode switching that is implementable not just for microwaves, but readily applicable also to light, acoustic or matter waves.

**Methods:**

**Numerical simulations.** In our numerical simulations we solve the Helmholtz equation Eq. (2) on a finite-difference grid by means of a Green's function method[30]. The transmission (reflection) amplitudes $t_{nm}$ ($r_{nm}$) are then determined by projecting the system's Green's function onto the flux-carrying modes in the semi-infinite leads which are attached to the scattering geometry. The corresponding intensities are given by $T_{nm} = |t_{nm}|^2$ and $R_{nm} = |r_{nm}|^2$, respectively. We choose the real part of the potential $V(x, y)$ to be finite (infinite) inside (outside) the cavity, corresponding to Dirichlet boundary conditions, and the imaginary part of the potential is determined such as to satisfy the protocol described in the main text.

**Experimental setup.** The experimental device is an aluminium waveguide with dimensions L × W × H = 2.38m × 5cm × 8mm. Fig. 3a shows the surface modulation that steers the modes around the EP. Our microwave experiment allows us to define the corresponding boundary conditions very accurately and to place the magnetized absorbing foam material (type: LS-10211 from ARC Technologies, W × H = 1.2mm × 5mm) with sub-wavelength (<0.5mm) precision. Additional absorbers (types: LS-14 and LS-16 from EMERSON&CUMING, W × L = 5cm × 17.5cm) are employed to mimic semi-infinite leads.

**Microwave measurements.** In order to probe the sinusoidal modes formed by the z-component of the electric field $E_z$[31], we use two microwave antennas 1.5m apart from each other. The antennas are fixed onto motor-controlled, moveable slides and measure the complex transmission signal outside of the modulated surface area at 2×2 points along the y-axis of the antennas. For the measurements we employ microwaves with a frequency around $\nu = 7.8\text{GHz}$, which is well below the cutoff frequency for $TM_0$ modes ($\nu_c = c/2H = 18.75\text{GHz}$), such that only the first two sinusoidal $TM_0$ modes contribute to the transport. By applying the two-fold sine transformation

$$t_{nm} = \sum_{y_1 y_2} t'(y_1, y_2) \sin\left(\frac{n\pi}{W} y_1\right) \sin\left(\frac{m\pi}{W} y_2\right),$$

where $t'(y_1, y_2)$ denotes the normalized transmission measured between antenna 1 at position $y_1$ and antenna 2 at position $y_2$, we obtain the transmission matrix $t_{nm}$ in its mode representation. The normalization is necessary to overcome the frequency dependent coupling of the two antennas, and is given by

$$t'(y_1, y_2) = \frac{t(y_1, y_2)}{\sqrt{(1-|\langle R_1^a \rangle|^2)(1-|\langle R_2^a \rangle|^2)}}.$$

Here, $t(y_1, y_2)$ describes the transmission and $\langle R_1^a \rangle$ ($\langle R_2^a \rangle$) denotes the measured reflection at antenna 1 (2), averaged over all positions $y_1$ ($y_2$) and a frequency window of 0.076GHz. The measured reflections are dominated by an imperfect impedance matching between the antenna and the channel. This results in a strong reflection signal originating from the antenna itself, which contains no information about the waveguide. However, dividing by $1 - \langle R^a \rangle$ normalizes the intensity fed into the system appropriately and is necessary to be able to compare the transmission in a broader frequency range.

**Acknowledgements** J.D., A.G. and S.R. are supported by the Austrian Science Fund (FWF) through projects no. SFB IR-ON F25-14, no. SFB-NextLite F49-P10 and no. I 1142- N27 (GePartWave). The computational results presented have been achieved in part using the Vienna Scientific Cluster (VSC). A.A.M. is supported by the National Council for Scientific and Technological Development (CNPq) grant no. 305519/2012-3. J.B. and U.K. acknowledge ANR project no. I 1142-N27 (GePartWave). F.L. acknowledges support by the Austrian Science Fund (FWF) through SFB-F41 VI-COM. T.J.M and P.R. are supported by the Austrian Science Fund (FWF) through DK CoQuS W 1210, SFB FOQUS F40, START grant Y 591-N16, and project OPSOQI (316607) of the WWTF. N.M. acknowledges I-Core: the Israeli Excellence Center "Circle of Light" and the Israel Science Foundation grants no. 298/11 and no. 1530/15 for their financial support.

Correspondence and requests for materials should be addressed to S.R. (stefan.rotter@tuwien.ac.at).

**Figures:**

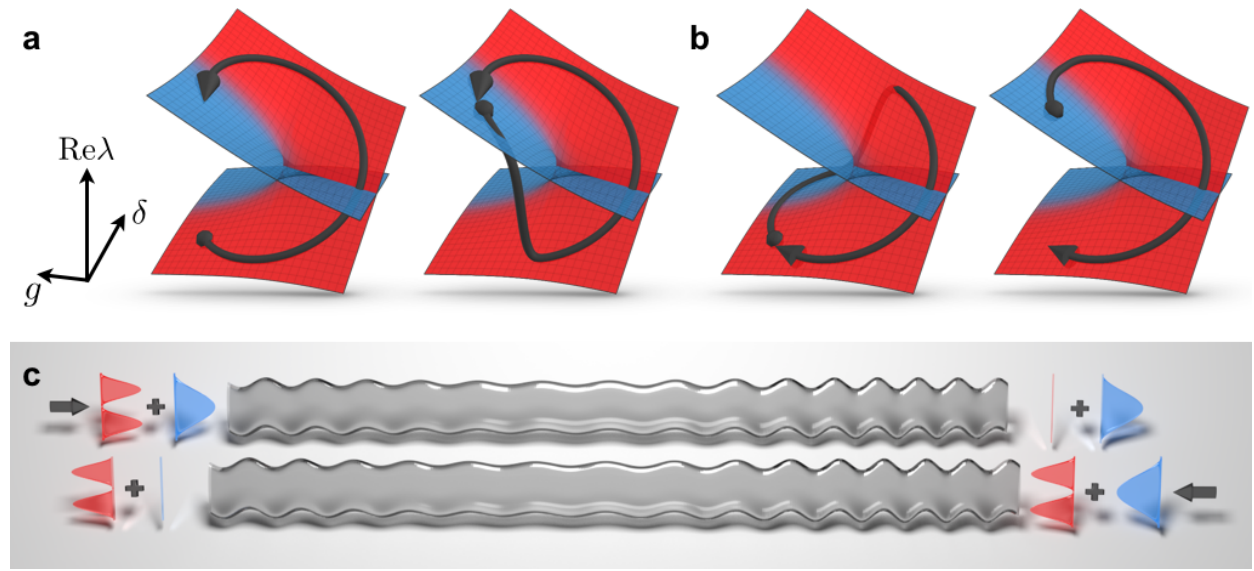

**Fig. 1. Mode evolution in the vicinity of an exceptional point.** To demonstrate the non-adiabatic nature of dynamically encircling an EP degeneracy, we show trajectories with different encircling directions, starting on both of the involved Riemann sheets (shown as red and blue surfaces). The results for the state evolution of the Schrödinger-type equation $i\partial_x \psi(x) = H(x)\psi(x)$, projected onto their respective Riemann sheets are shown as black lines: the larger the contribution of an eigenvector, the closer it follows the corresponding eigensheet[21]. **a**, Dynamics of two states with starting points on different sheets during a counter-clockwise loop around the EP (as seen from the top). **b**, Same as **a** for a clockwise loop. In both **a** and **b** the end

points of the loops depend only on the encircling direction, not on their starting point. **c**, Schematic of an asymmetric mode-switch that projects the above EP-encircling to a waveguide that strongly attenuates one of its two transverse modes depending on the injection direction. The parameter-space trajectories describing counter-clockwise and clockwise loops around the EP shown in panels **a** and **b** correspond to left and right injection, respectively.

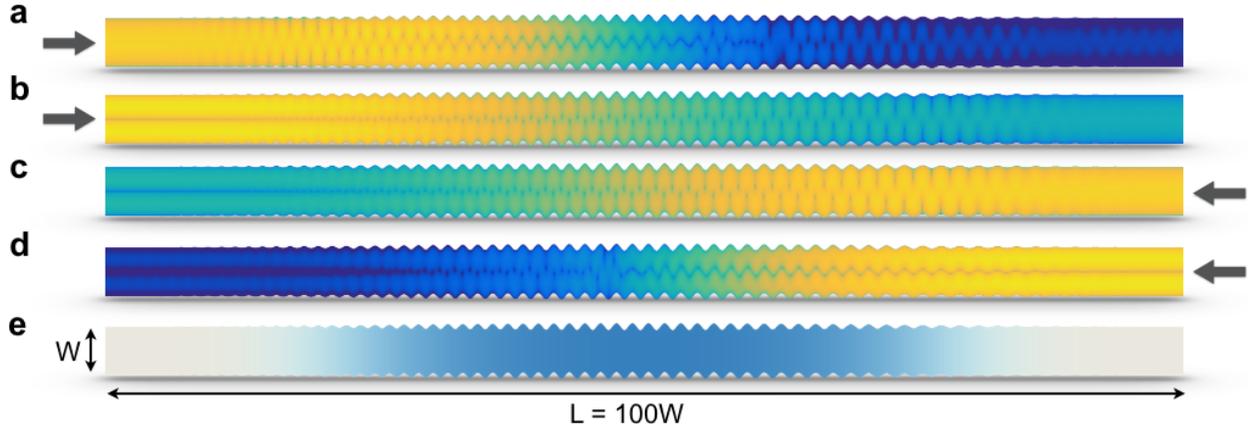

**Fig. 2. Chiral transport in the presence of bulk absorption.** Numerically simulated modal wavefunction intensities for a waveguide with a length-to-width ratio of $L/W = 100$ (the depicted dimensions are not to scale). Shown are results for different input modes and injection directions: Arrows indicate the side from which the waveguide is excited, the first mode is injected in panels **a** and **c**, the second mode in **b** and **d**, respectively. Note that we employ a logarithmic scale for the respective intensities since the overall dissipation is very strong as evident from the corresponding values for transmission $T_{nm}$ from mode n into mode m: $T_{11} = 2.5 \cdot 10^{-11}$, $T_{21} = 8.9 \cdot 10^{-7}$, $T_{12} = 7.0 \cdot 10^{-14}$ and $T_{22} = 8.4 \cdot 10^{-12}$. The normalized mode profiles at the waveguide exit, which clearly show the efficient mode-switching, are shown in Supplementary Fig. 7. **e**, Plot of the absorption strength that is gradually switched on-and-off, but uniform in transverse direction. The parameters used to describe a loop around the EP determine the waveguide boundary modulation: Its amplitude $\sigma$ is proportional to the coupling $g$, the oscillation frequency $k_b$ in turn is controlled by the detuning $\delta$ (see Supplementary Fig. 8 for an example that displays the actual $L/W$ dimensions and where the change in detuning is readily visible). Specifically, for the above waveguide, $\omega W/c\pi = 2.05$, $\sigma_0/W = 0.07$, $\delta_0 W = 0.5$ and

$\rho W = -0.5$.

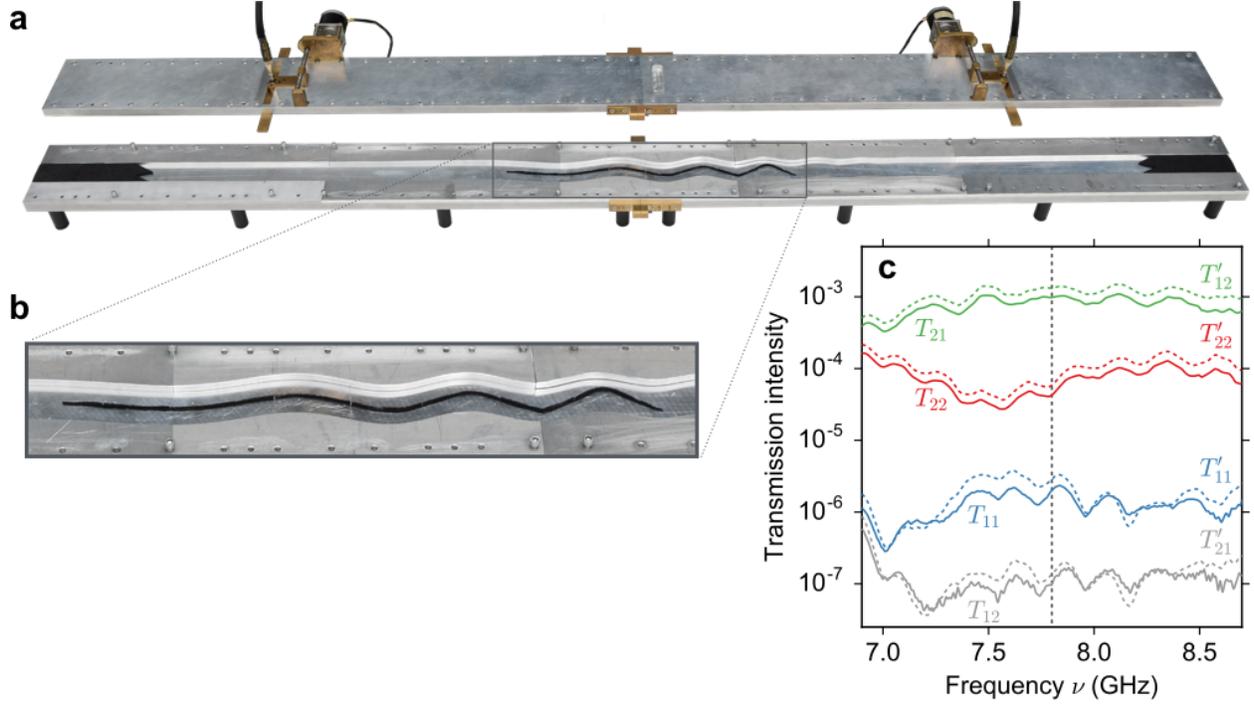

**Fig. 3. Microwave measurements. a**, Photograph of the optimized waveguide channel used in the experiment, with a surface modulated region of length $L = 1.25$ m and width $W = 5$ cm. Within this setup, input and output antennas are placed 1.5m apart (shown on the top plate). Black foam is used both as an absorber in the center of the waveguide (magnified in panel **b**) and to mitigate the reflection into entrance and exit leads. The setup is engineered for a target frequency of $\nu = 7.8$ GHz (shown by a dashed vertical line in **c**), but the design ensures applicability over a broad frequency interval. **c**, Measured frequency dependent transmission intensities $T_{nm}(T'_{nm})$ from mode $n$ into mode $m$ for injection from left (right) are shown by solid (dashed) lines. The waveguide parameters here are the following: $\omega W/c\pi = 2.6$, $\sigma_0/W = 0.16$, $\delta_0 W = 1.25$ and $\rho W = -1.8$.